\documentclass[%
 reprint,%
 amssymb, amsmath,%
aps 
]{revtex4-1}

\usepackage{graphicx}
\usepackage{dcolumn}
\usepackage{bm}

\usepackage{graphicx}
\usepackage{amsmath}
\usepackage{amsfonts}
\usepackage{latexsym}
\usepackage{amsbsy}
\usepackage{epstopdf}
\usepackage{bm}%

\newcommand{\ket}[1]{\vert #1 \rangle}

\usepackage{color}

\usepackage{ulem} 
\begin{document}



\title{Continuous radio frequency electric-field detection through\\adjacent Rydberg resonance tuning}

\author{Matthew T. Simons}
\email{matthew.simons@nist.gov}
\author{Alexandra B. Artusio-Glimpse}
\author{Christopher L. Holloway}
\affiliation{National Institute of Standards and Technology, Boulder,~CO~80305, USA}
\author{Eric Imhof}
\author{Steven R. Jefferts}
\affiliation{Northrop Grumman, Woodland Hills, CA, USA}
\author{Robert Wyllie}
\author{Brian C. Sawyer}
\affiliation{Georgia Tech Research Institute, Atlanta, Georgia 30332, USA}
\author{Thad G Walker}
\affiliation{Dept. of Physics,  University of Wisconsin-Madison, Madison, WI 53706}

\date{\today}

\begin{abstract}
We demonstrate the use of multiple atomic-level Rydberg-atom schemes for continuous frequency detection of radio frequency (RF) fields. Resonant detection of RF fields by electromagnetically-induced transparency and Autler-Townes (AT) in Rydberg atoms is typically limited to frequencies within the narrow bandwidth of a Rydberg transition. By applying a second field resonant with an adjacent Rydberg transition, far-detuned fields can be detected through a two-photon resonance AT splitting. This two-photon AT splitting method is several orders of magnitude more sensitive than off-resonant detection using the Stark shift. We present the results of various experimental configurations and a theoretical analysis to illustrate the effectiveness of this multiple level scheme. These results show that this approach allows for the detection of frequencies in continuous band between resonances with adjacent Rydberg states.

\end{abstract}

\maketitle


\section{\label{sec:intro}Introduction}

Rydberg atoms have been increasingly investigated as radio frequency (RF) field sensors~\cite{osterwalder1999, holloway2014tantprop, fan2015josab,anderson2016prapp,cox2018prl,meyer2018apl,song2019oe,meyer2020jphysb,simons2019apl,holloway2019antwireless}. Rydberg atoms can detect microwave fields through both a Stark shift due to their polarizability~\cite{gallagher1994book} and Autler-Townes (AT) splitting~\cite{AutlerTownesPR100,tanasittikosol2011jphysb,sedlacek2012nature} from resonant transitions. Resonant interactions are much more sensitive than the Stark effect, though they are limited to detection of discrete frequencies. However, Rydberg states can have strong couplings to many nearby states~\cite{robinson2021pra}. We show that by coupling multiple Rydberg levels we can effectively extend the frequency range of the resonant interaction, increasing the sensitivity to microwave fields far-detuned from a resonance transition. 
This paper discusses the architecture of this multi-level Rydberg system and nuances of tuning sensitivity to off-resonant signals using AT splitting. 

\begin{figure}[t]
\includegraphics[width = \columnwidth]{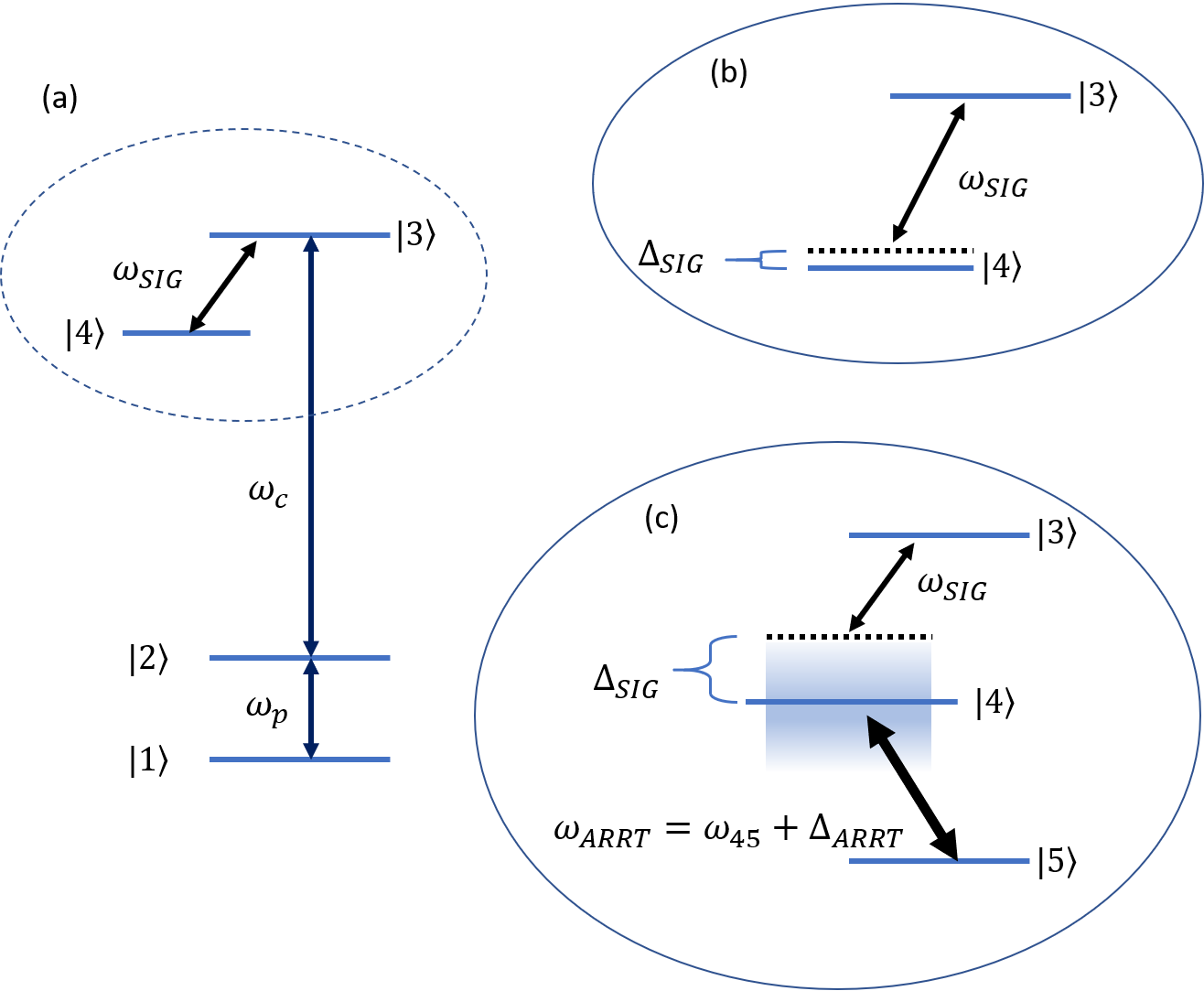}
\caption{\label{fig:1} (a) Typical 4-level EIT/AT scheme where $\omega_p$ and $\omega_c$ are the optical probe and pump fields and $\omega_{SIG}$ is the RF signal that causes AT splitting for detection. (b) The RF field can be detuned over a small range from the $\ket{3} - \ket{4}$ transition and still cause AT splitting. (c) This range can be extended by coupling to an adjacent resonance, $\ket{4} - \ket{5}$.}
\end{figure}

The typical Rydberg atom-based RF electric (E) field detection technique relies on a 4-level electromagnetically-induced transparency (EIT) scheme, as shown in Fig.~\ref{fig:1}(a). This scheme includes a ground-state probe laser (which couples levels $\ket{1}$ and $\ket{2}$), a coupling laser that populates a Rydberg state ($\ket{3}$, coupled from level $\ket{2}$), and the RF field of interest (signal field, or SIG) which couples two Rydberg states (levels $\ket{3}$ and $\ket{4}$). The addition of the resonant RF field (labeled as $\omega_{SIG}$) results in AT splitting. The AT split gives a direct SI-traceable measurement of the SIG RF E-field strength \cite{sedlacek2012nature, holloway2014tantprop, fan2015josab, holloway2017transec}.

Using standard EIT/AT techniques, E-field strength can be routinely measured down to a few mV/cm \cite{sedlacek2012nature, holloway2014tantprop, holloway2017transec, holloway2014apl}. With optical homodyne \cite{kumar2017scireports} or RF heterodyne (a Rydberg atom-based mixer) \cite{gordon2019aipadvances, jing2020nphys} techniques with EIT, E-field strengths down to 55-700~nV/cm$\sqrt{Hz}$ have been measured. While the standard resonant EIT approach is capable of weak field detection, it is limited by the bandwidth of the EIT, on the order of $\pm 100$~MHz. While this bandwidth depends partially on the Rydberg state lifetimes, these do not vary much for the Rydberg states typically used for electrometry. 

\begin{figure}[ht]
\includegraphics[width = 0.75\columnwidth]{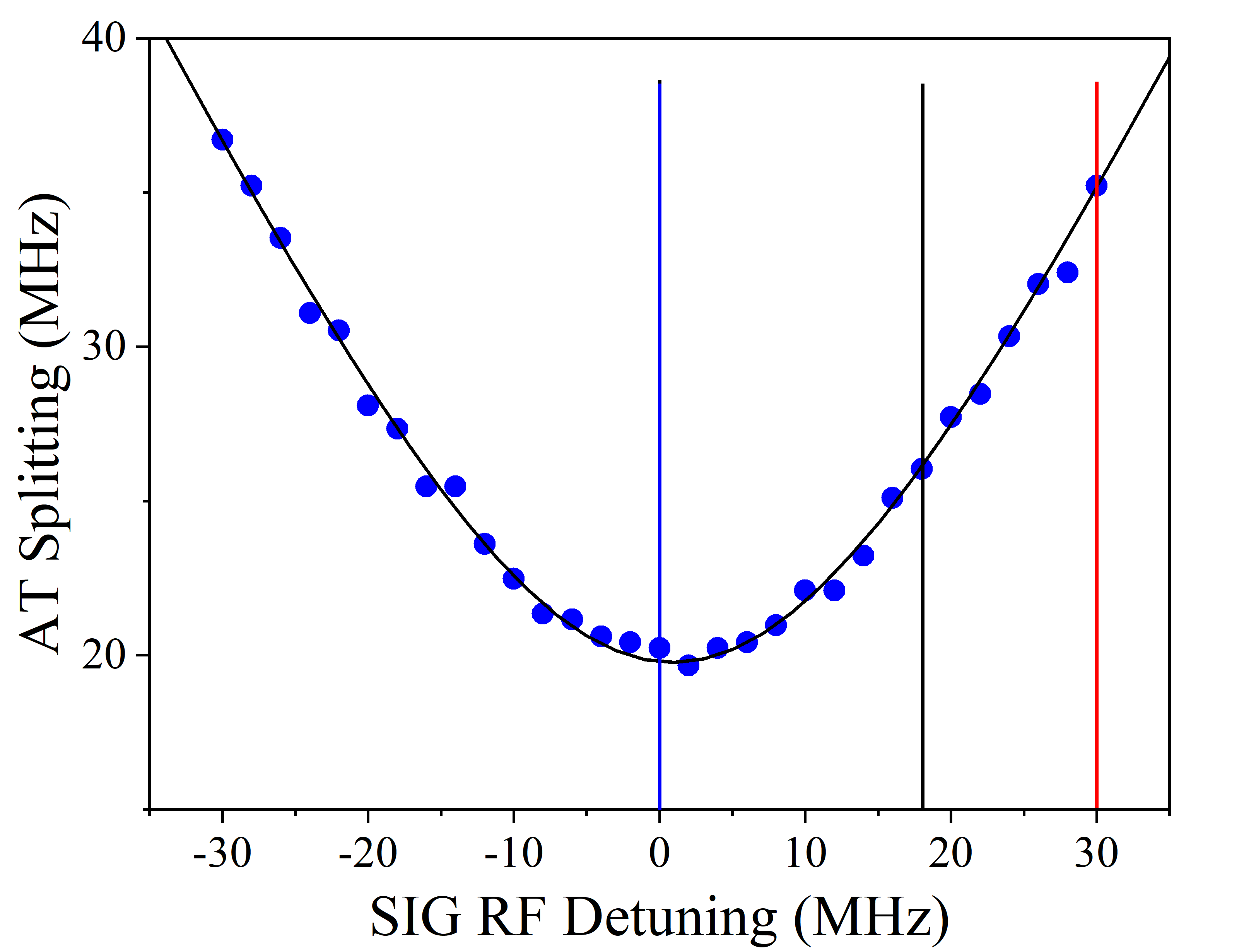}\\
\vspace*{12pt}
\includegraphics[width = 0.75\columnwidth]{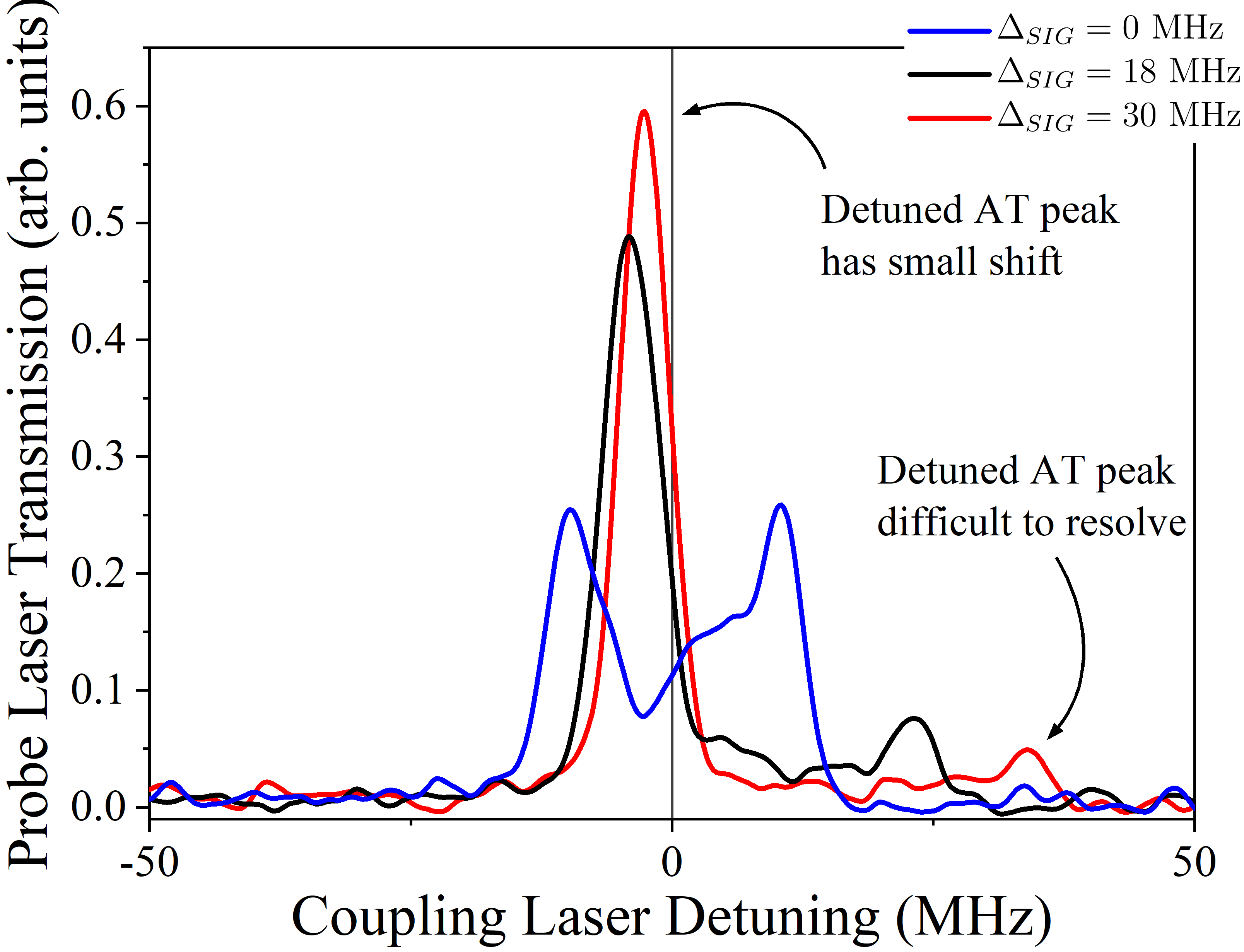}
\caption{\label{fig:2} RF detuning: (top) AT splitting as a function of RF detuning and (bottom) EIT signal for several RF detunings. As the detuning increases, one AT peak shifts further away and reduces in amplitude, while the other shifts back towards the zero field case.}
\end{figure}

When an RF field is detuned from its resonant RF transition frequency there are two significant effects on the observed AT splitting of the EIT signal, which are discussed in detail in Ref. \cite{simons2016apl}. The AT splitting as a function of SIG detuning is shown in Fig.~\ref{fig:2}(a). For illustration, we reproduced the AT spectra at three values of SIG detuning in Fig.~\ref{fig:2}(b). First, the two peaks of the EIT signal become asymmetric (i.e., the heights of the two peaks are not the same). The second effect of SIG detuning is that the separation between the two peaks increases with RF detuning, with a reduction in the amplitude of one peak as it spreads further while the other peak returns to its unperturbed frequency and height. Once the frequency of the SIG is detuning far enough, the AT splitting can no longer be observed.  

Thus, the standard four-level EIT approach is effectively only sensitive at discrete RF frequencies, given that, for a given Rydberg state (or principal quantum number $n$) an RF field can only be measured within a limited band around the Rydberg atomic transition frequency. This limits the practical uses of a resonant Rydberg atom-based RF field sensor. When the SIG RF field is far off-resonance, it is possible to detect through the AC Stark shift of the three-level EIT. While the Stark shift approach allows of the detection of the RF field over a wide range of frequencies, it is much less sensitive than the resonant AT approach, as large field levels are required to observe these Stark shifts. For this technology to progress to a deployable device, a method for sensitive measurements over a continuous frequency range is required.

In order to change the resonant transition frequency, a second field can be applied to shift one of the Rydberg energy levels, as shown in Fig~\ref{fig:1}(b). This scheme uses an additional RF field coupled to an adjacent Rydberg transition. The first four levels of our five-level scheme use the same ladder schemes described above, while the fifth  ($\ket{5}$) corresponds to the adjacent Rydberg state. A similar five-level configuration was investigated previously, where interesting atomic spectra that are not accessible with the basic four-level system were observed \cite{robinson2021pra}. In this paper we show that the addition of the adjacent Rydberg resonance tuning (ARRT) RF field allows for EIT/AT detection of continuous frequencies of the SIG field.  We show that this continuous frequency detection can be achieved in two ways: either by varying the strength or the frequency of the ARRT field.

\section{\label{sec:setup}Experimental Setup}

\begin{figure}[ht]
\scalebox{.95}{\includegraphics[width = \columnwidth]{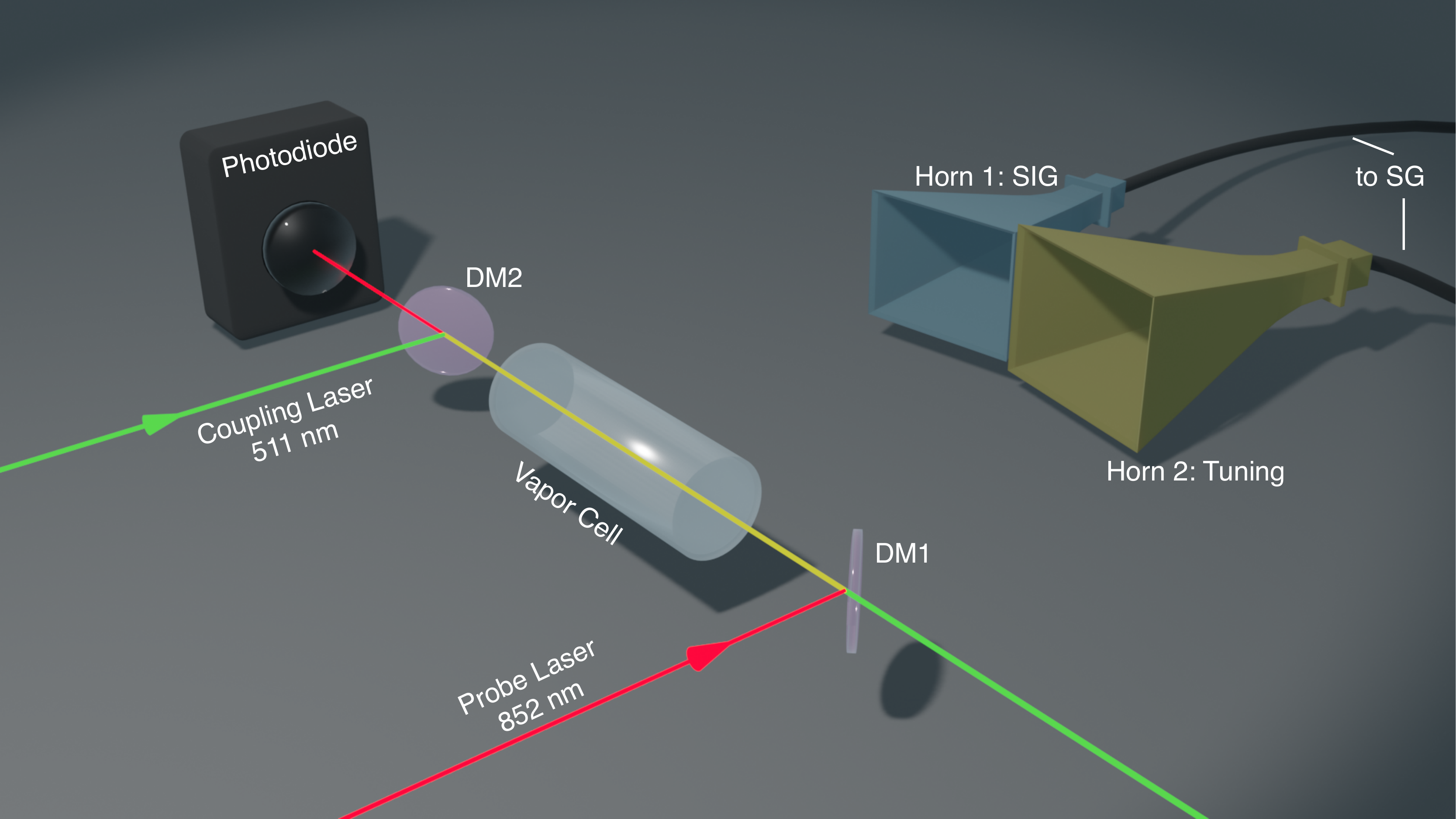}}
\caption{\label{fig:setup} Experimental setup of the five-level EIT scheme. SG: signal generator, DM: dichroic mirror.}
\end{figure}

In these experiments we use cesium ($^{133}$Cs) placed in a cylindrical glass vapor cell with diameter of $8$~mm and a length of $30$~mm.  The five-level system consists of the $^{133}$Cs $6~S_{1/2}$ ground state $\ket{1}$, the $6~P_{3/2}$ excited state $\ket{2}$, and three Rydberg states $\ket{3}$, $\ket{4}$, $\ket{5}$. The first Rydberg state $\ket{3}$ is the $68~S_{1/2}$, coupled to $\ket{2}$ with a $509~$nm coupling laser. The other two Rydberg states $\ket{4}$ and $\ket{5}$ are varied for the experiments to detect various SIG RF frequencies.
The experimental setup is depicted in Fig.~\ref{fig:setup}, which consists of an $850~$nm probe laser, a $509~$nm coupling laser, two RF signal generators (SG), two horn antennas, and a photodetector connected to an oscilloscope.

The probe laser is locked to the D2 transition $6~S_{1/2}(F=4) \rightarrow 6~P_{3/2}(F=5)$ with a wavelength of $\lambda_p=852.35$~nm \cite{SteckCsData}.  To produce an EIT signal, we apply a counter-propagating coupling laser with $\lambda_c \approx 508.98$~nm and scan it across the $6~P_{3/2} \rightarrow 68~S{1/2}$ Rydberg transition. We use a lock-in amplifier to enhance the EIT signal-to-noise ratio by modulating the coupling laser amplitude with a $37~$kHz square wave with a $50~\%$ duty cycle. This removes the background and isolates the EIT signal. RF sources are used to couple to the Rydberg states  $\ket{3}$, $\ket{4}$ and $\ket{5}$. To generate these RF fields, the output of two signal generators (SG) are connected to two different horn antennas (referred to as Horn~1 and Horn~2).
Horn~1 and Horn~2 were placed 23~cm and 21~cm from the laser beam locations in the vapor cell, respectively. Horn~1 produces the SIG field coupling Rydberg states $\ket{3}$ and $\ket{4}$. Horn 2 produces the ARRT field used to couple Rydberg states $\ket{4}$ and $\ket{5}$.
In these experiments, the optical beams and the RF electric fields are co-linearly polarized, in that the E-field vectors are all pointing in the same direction. In our case, the E-field vectors are pointing up from the optics table, perpendicular to the plane that contains all of the propagation vectors.

\section{\label{sec:results}Experimental Results}

Continuous frequency detection for the SIG field (which couples states $\ket{3}$ and $\ket{4}$) can be achieved either by varying the Rabi frequency of the ARRT field or detuning the ARRT field from resonance with the $\ket{4} \rightarrow \ket{5}$ transition. In this section we investigate both approaches.

\begin{figure}[hb]
\includegraphics[width = \columnwidth]{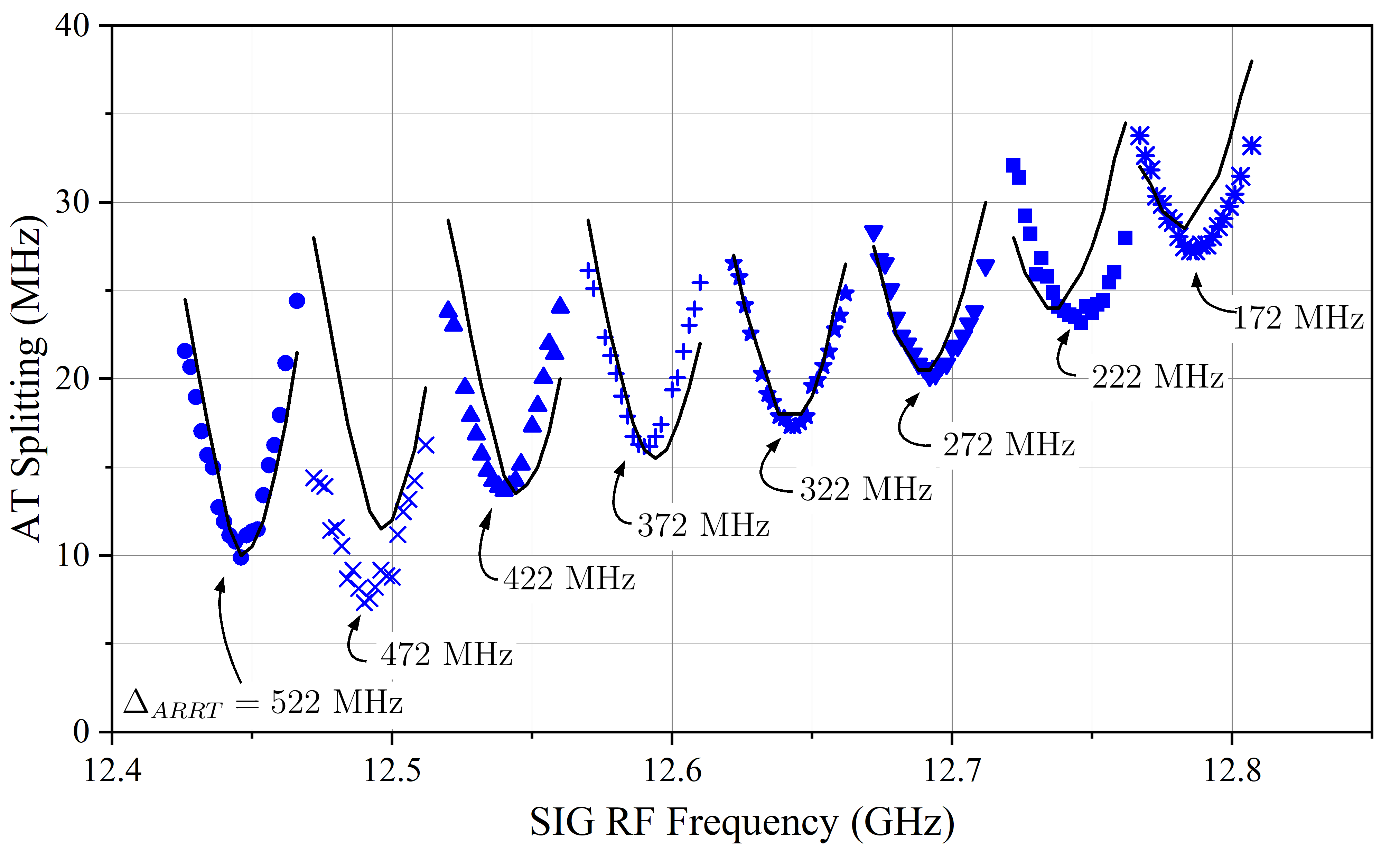}
\caption{\label{fig:alldetuning} SIG detuning $\delta_{SIG}$ for various values of $\Delta_{ARRT}$. A description of the numerical model used to generate the simulated results (solid lines) is given in Section~\ref{sec:model}.}
\end{figure}

\begin{figure}[hb]
\scalebox{.27}{\includegraphics{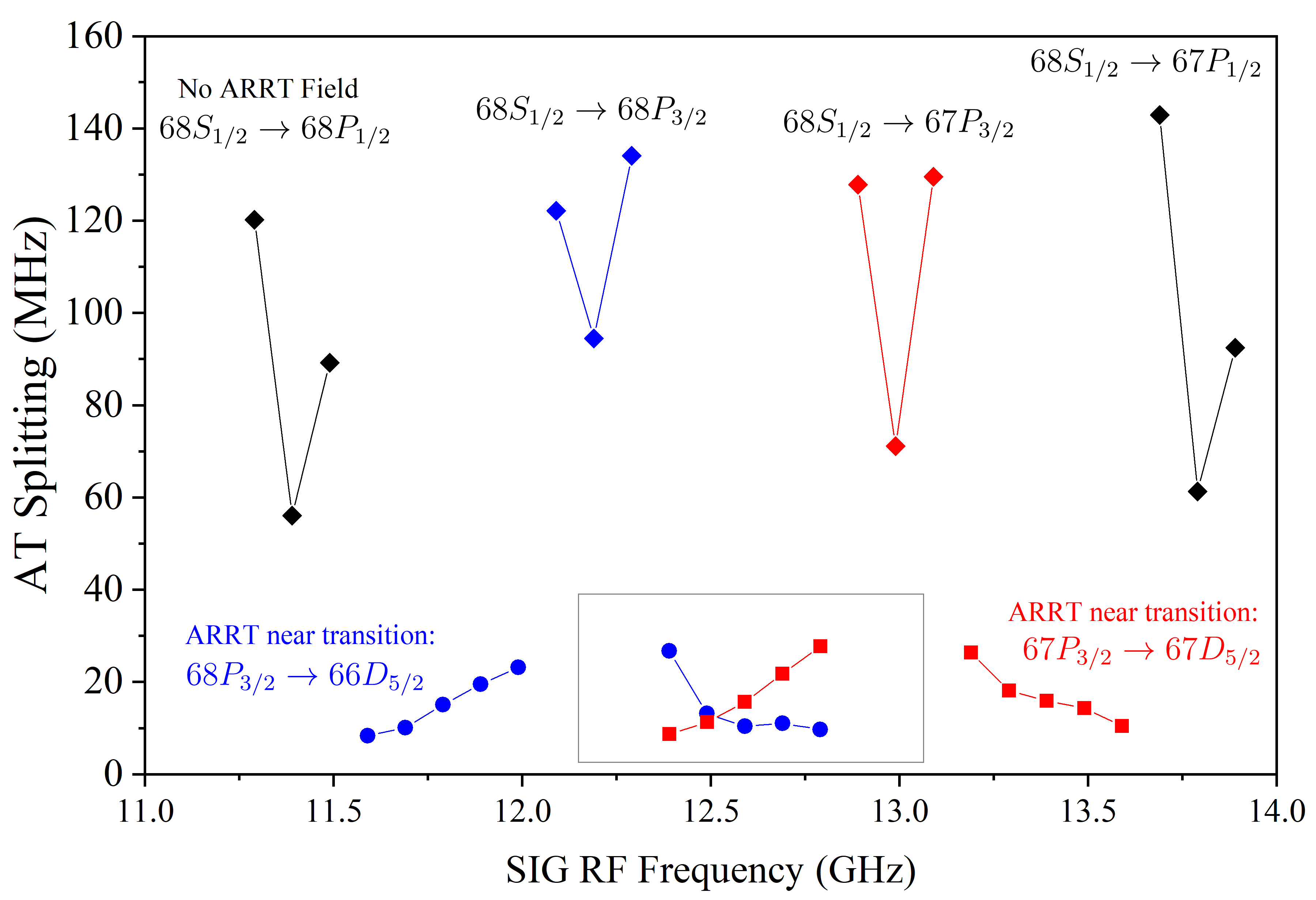}}
\caption{\label{several-n} Measurements of the AT splitting by a SIG field from both direct resonances (diamonds) and with an ARRT field (squares and circles). The blue circles are AT splittings enabled by applying an ARRT field near the $68~P_{3/2} - 66~D_{5/2}$ transition to shift the $68 P_{3/2}$ state, and the red squares were made using an ARRT field near the   $67~P_{3/2} - 65~D_{5/2}$ transition to shift the $67~P_{3/2}$ state. This shows that the application of an appropriate adjacent resonance field can enable resonant detection over a continuous frequency range spanning several Rydberg states.}
\end{figure}

\subsection{Detuning the Frequency of the ARRT field}

When the ARRT field is off, the measurements of AT-splitting versus the SIG detuning ($\Delta_{SIG}$) are shown in Fig.~\ref{fig:2}(a), for a given Rabi frequency. The asymmetric amplitudes of the peaks illustrate that without the ARRT field, the SIG can be detected only when $\omega_{SIG}$ is within $\pm50~$MHz of the transition frequency for the $68~S_{1/2} \leftrightarrow 67~P_{3/2}$ states. 

\begin{figure}[htbp]
\includegraphics[width = \columnwidth]{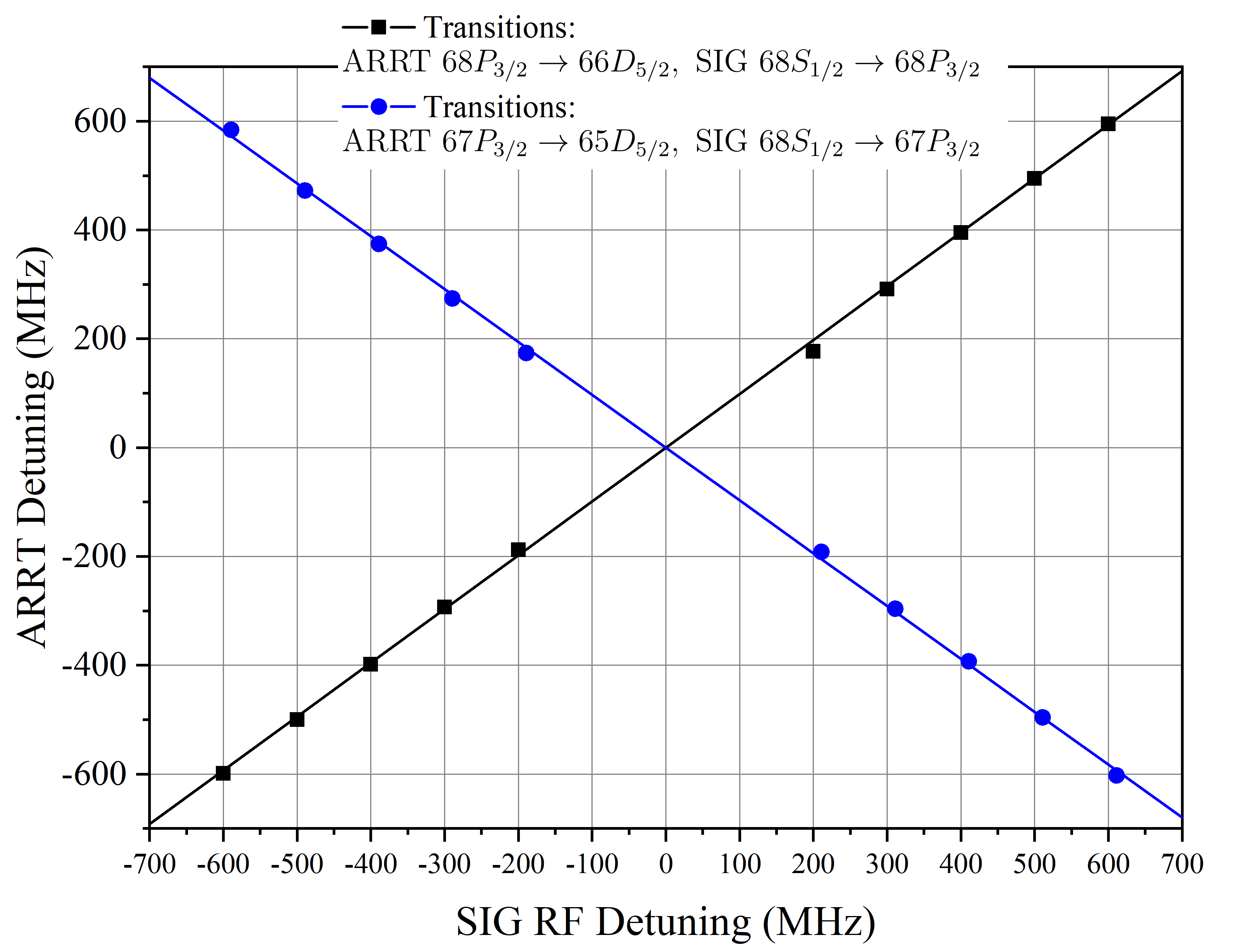}
\caption{\label{linear} ARRT detunings that give symmetric AT splitting versus SIG detuning. The $68~P_{3/2}$ is higher in energy than the $68~S_{1/2}$ state, so there is a linear relationship in the two-photon detuning, while the $67~P_{3/2}$ is lower in than the $68~S_{1/2}$ state, so the two-photon detuning has an inverse relationship.}
\end{figure}

However, by applying an ARRT field to shift the Rydberg state, the frequency range for which a SIG field can be detected dramatically increases, as illustrated in Fig.~\ref{fig:alldetuning}. Here we plot several $\Delta_{SIG}$ curves [similar to the one given in Fig.~\ref{fig:2}(a)] for different values of the ARRT field detuning ($\Delta_{ARRT}$).  In these data, $\ket{4} \leftrightarrow \ket{5}$ corresponds to $67~P_{3/2} - 65~D_{5/2}$ with frequency $\omega_{45} = 23.89~$GHz.  In fact, by varying $\Delta_{ARRT}$, it is possible to detect a signal over a continuous frequency that spans from one principal quantum number {\it n} (for $\ket{3} \leftrightarrow \ket{4}$) to another by simply changing $\Delta_{ARRT}$. This is shown in Fig.~\ref{several-n} where we show continuous frequency detection from $11.25$ to $13.9~$GHz by measuring an AT splitting either directly or enabled by an ARRT field. Keeping $\Omega_{ARRT}$ constant and relatively low, the optimal detection of a particular $\Delta_{SIG}$ occurs for $\Delta_{ARRT} = \pm \Delta_{SIG}$, as shown in Fig.~\ref{linear}. In these results we see that if we use $\ket{3} = 68 S_{1/2}$, $\ket{4} = 68 P_{3/2}$, and $\ket{5}= 66~D_{5/2}$, an increase in $\Delta_{ARRT}$ corresponds to a linear increase in the detectable frequency detuning for SIG. When we switch to $\ket{4} = 67~P_{3/2}$ and $\ket{5} = 65~D_{5/2}$, a decrease in $\Delta_{ARRT}$ corresponds to a linear decrease in the detectable frequency detuning for SIG.
This approach can be used to measure continuous frequencies for the SIG field over several Rydberg states. This is shown in Fig.~\ref{several-n}.

\subsection{Power Tuning of the ARRT field}

\begin{figure}[htbp]
\includegraphics[width = \columnwidth]{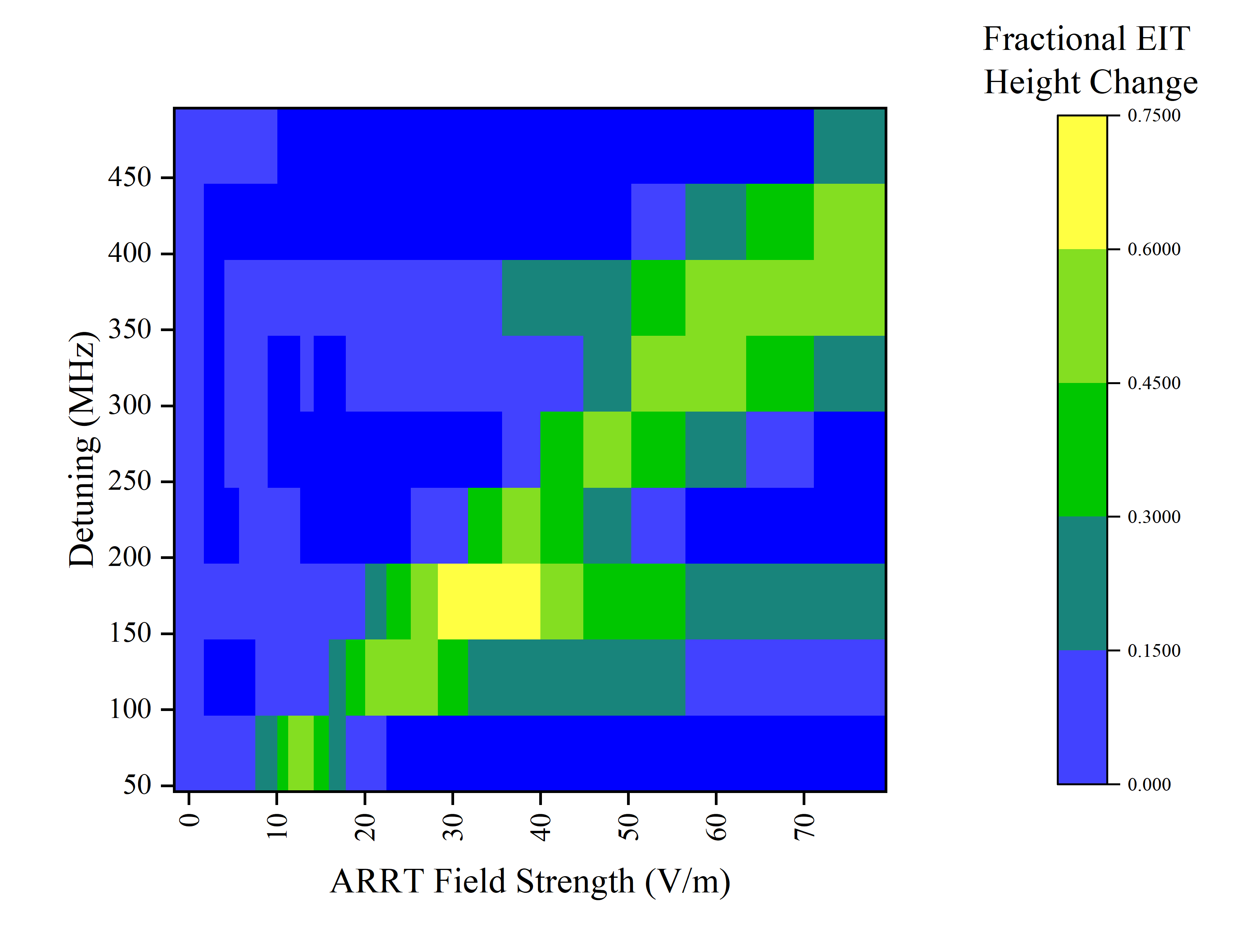}
\caption{\label{fig:pow1} Fractional reduction in EIT amplitude (change in EIT amplitude over the unperturbed amplitude) versus ARRT field Rabi frequency $\Omega_{ARRT}$ and SIG field detunings $\Delta_{SIG}$. The maximum EIT amplitude reduction for a given $\Delta_{SIG}$ occurs at a larger $\Omega_{ARRT}$ as the detuning increases.}
\end{figure}

\begin{figure}[htbp]
\includegraphics[width = \columnwidth]{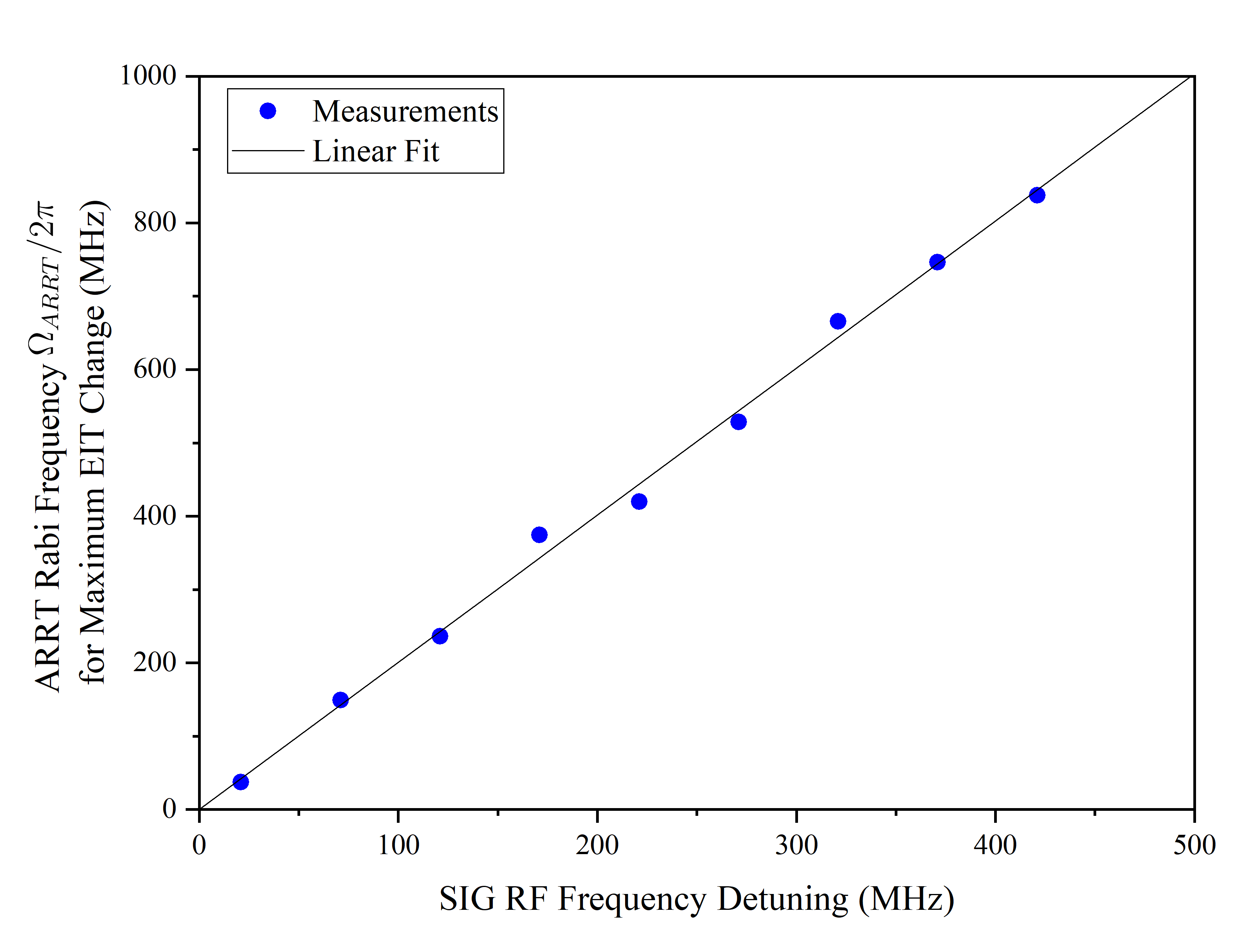}
\caption{\label{fig:pow2} The ARRT field Rabi frequency $\Omega_{ARRT}$ for maximum sensitivity increases linearly with SIG field detunings $\Delta_{SIG}$.}
\end{figure}

The two-photon response can also be achieved by varying the Rabi frequency of the ARRT field, $\Omega_{ARRT}$. In this case we leave the frequency of the ARRT field resonant with the tuning transition ($\Delta_{ARRT} = 0$), and vary the power delivered to the ARRT field horn antenna in order to maximize the EIT response for a given $\Delta_{SIG}$. To measure this effect, we work with weak SIG strengths such that they do not cause AT splitting, but do reduce the height of the EIT peak.
Fig.~\ref{fig:pow1} shows the reduction in EIT height (where a positive value reflects a decrease in the EIT height) versus $\Omega_{ARRT}$, for nine different $\Delta_{SIG}$. As the SIG field is detuned away from resonance, the ARRT field strength must be increased in order to see an optimal two-photon coupling between the three Rydberg levels. Fig.~\ref{fig:pow2} shows that the optimal ARRT field strength varies linearly with $\Delta_{SIG}$. 

\begin{figure}[htbp]
\includegraphics[width = \columnwidth]{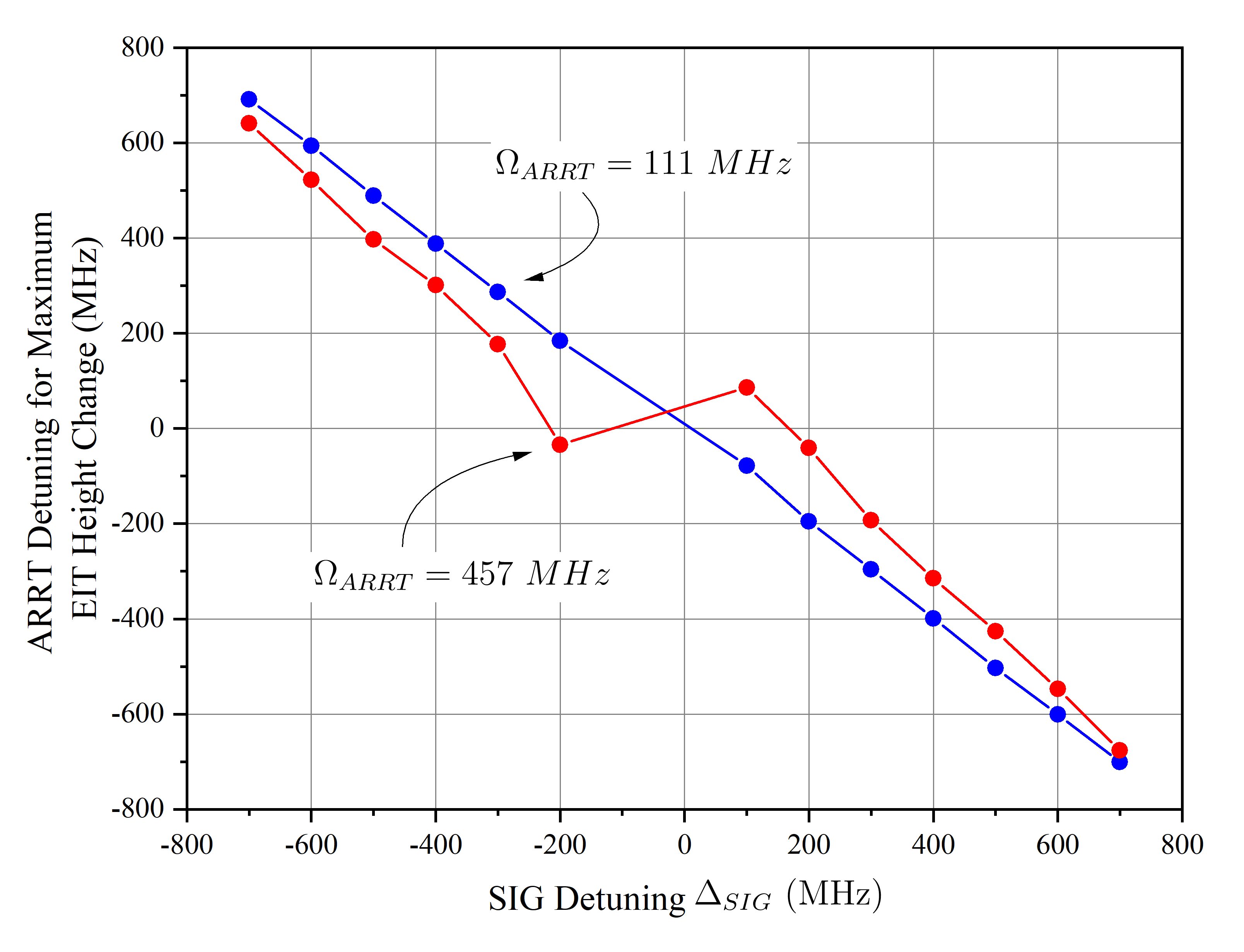}
\caption{\label{fig:pow3} ARRT field detuning $\delta_{ARRT}$ is varied to maximize response to various SIG field detunings $\delta_{SIG}$, for two ARRT field strengths. For a weak ARRT field, optimal $\delta_{ARRT}$ varies linearly with the detuning of the SIG field. For a strong ARRT field, optimal $\delta_{ARRT}$ is reduced, and becomes nonlinear. The asymmetry is likely due to an AC stark shift caused by the strong ARRT field.}
\end{figure}

We can also vary both field strength and frequency of the ARRT field to maximize sensitivity to SIG. Fig.~\ref{fig:pow3} shows how the ARRT field Rabi frequency affects the optimal ARRT field detuning $\Delta_{ARRT}$. For weak $\Omega_{ARRT}$ the optimal $\Delta_{ARRT}$ varies linearly with $\Delta_{SIG}$, especially far from the on-resonance condition for the SIG field, as shown in the previous section. However, for a strong $\Omega_{ARRT}$ the optimal $\Delta_{ARRT}$ is shifted lower and becomes increasingly nonlinear near the $\Delta_{SIG} = 0$ resonance. 

\subsection{Uncertainties}
A discussion of the contributions to the uncertainties of the Autler-Townes splitting measurement can be found in \cite{sedlacek2012nature} and \cite{simons2018emceurope}. Measurements of the AT splitting typically have an uncertainty of $\sim 1~\%$, arising from the uncertainty in peak fitting and in the frequency scale. There can be large uncertainty arising from the perturbation of the RF field due to the dielectric vapor cell, which affects the determination of the field strength from the AT measurement. In these experiments, there is uncertainty in the Rabi frequencies of the ARRT and the SIG due to the different standing wave patterns inside the vapor cell. This is one significant source of disagreement between the models and experiment. Further work will analyze uncertainty contributions for this ARRT method of field strength detection, in comparison to the contributions for the standard AT method.

\section{Theoretical Modeling}\label{sec:model}

\begin{figure}[t]
\centering
\scalebox{.4}{\includegraphics*{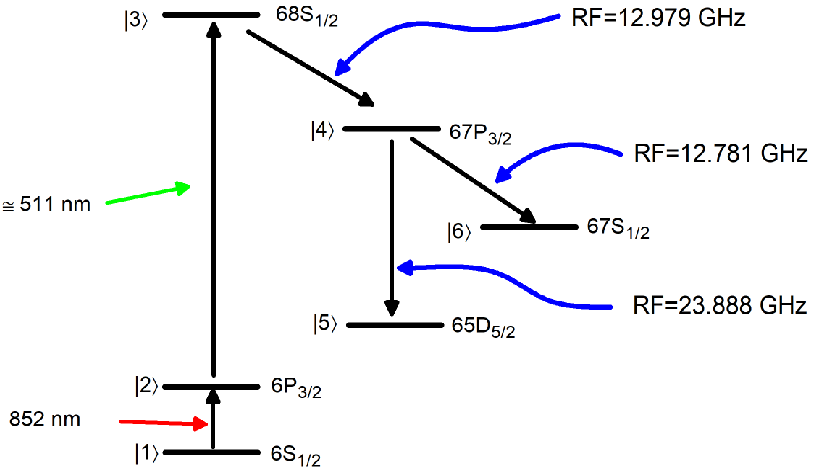}}\\
\caption{Six-level scheme used to model the experimental data.}
\label{multilevel}
\end{figure}

We present a theoretical model to depict the observed continuous frequency detection shown in the experimental results. Note that for the Rydberg states used in these experiments, the frequency of SIG field can simultaneously couple to two sets of states, $68~S_{1/2} - 67~P_{3/2}$ and $67~S_{1/2} - 67~P_{3/2}$. This is due to the fact that the difference between the transition frequencies of these two sets of states is 198~MHz (see Fig.~\ref{multilevel}) and these transitions have similar dipole moments ($d$). The radial part of $d$ for the two states are $d=4396.3\,e\,a_o$  for $68~S_{1/2} - 67~P_{3/2}$ and $d=4500.9\,e\,a_o$  for $67~S_{1/2} - 67~P_{3/2}$, where $e$ is the elementary charge (in units of C) and $a_0$ is the Bohr radius (in units of m).  Given that the SIG field couples to both these states, Rabi frequencies for the two transitions are related by the ratio of these two dipole moments and given by
\begin{equation}
  \Omega_{RF46} = 1.02\,\, \Omega_{RF34} = 1.02~\Omega_{SIG}
\end{equation}

To account for this potential simultaneous coupling, we model the six-level scheme shown in Fig.\ref{multilevel}. The model used here is similar to the one presented \cite{robinson2021pra, holloway2017japplphys}, where the main difference is the atomic species used.  In the model for the experimental results of the paper, we use $^{133}$Cs, while in Refs. \cite{robinson2021pra, holloway2017japplphys} $^{85}$Rb is used. For this six-level $^{133}$Cs model we follow a similar process to the one given in \cite{robinson2021pra, holloway2017japplphys} and we start by noting that the power of the probe beam measured on the detector (the EIT signal, i.e., the probe transmission through the vapor cell) is given by \cite{berman2011book}
\begin{equation}
P=P_0 \exp\left(-\frac{2\pi L \,\,{\rm Im}\left[\chi\right]}{\lambda_p}\right)=P_0 \exp\left(-\alpha L\right) \,\,\, ,
\label{intensity}
\end{equation}
where $P_0$ is the power of the probe beam at the input of the cell, $L$ is the length of the cell, $\lambda_p$ is the wavelength of the probe laser,  $\chi$ is the susceptibility of the medium seen by the probe laser, and $\alpha=2\pi{\rm Im}\left[\chi\right]/\lambda_p$ is Beer's absorption coefficient for the probe laser.  The susceptibility for the probe laser is related to the density matrix component ($\rho_{21}$)  by the following \cite{berman2011book}
\begin{equation}
\chi=\frac{2\,{\cal{N}}_0\wp_{12}}{E_p\epsilon_0} \rho_{21_D} =\frac{2\,{\cal{N}}_0}{\epsilon_0\hbar}\frac{(d\, e\, a_0)^2}{\Omega_p} \rho_{21_D}\,\,\, ,
\label{chi1}
\end{equation}
where $d=2.02$\cite{SteckCsData} is the normalized transition-dipole moment for the probe laser, $\Omega_p$ is the Rabi frequency for the probe laser in units of rad/s. The subscript $D$ on $\rho_{21}$ presents a Doppler averaged value. ${\cal{N}}_0$ is the total density of atoms in the cell and is given by
\begin{equation}
{\cal{N}}_0= \frac{p}{k_B T} \,\, ,
\label{nn}
\end{equation}
where $k_B$ is the Boltzmann constant, $T$ is temperature in Kelvin, and the pressure $p$ (in units of Pa) is given by \cite{SteckCsData}
\begin{equation}
p=10^{9.717-\frac{3999}{T}} 
\label{ppp}
\end{equation}
In eq. (\ref{chi1}), $\wp_{12}$ is the transition-dipole moment for the $\ket{1}$-$\ket{2}$ transition, $\epsilon_0$ is the vacuum permittivity, and $E_p$ is the amplitude of the probe laser E-field.

The density matrix component ($\rho_{21}$) is obtained from the master equation \cite{berman2011book}
\begin{equation}
\dot{\rho}=\frac{\partial \rho}{\partial t}=-\frac{i}{\hbar}\left[H,\rho\right]+{\cal{L}} \,\,\, ,
\label{me}
\end{equation}
where $H$ is the Hamiltonian of the atomic system under consideration and ${\cal{L}}$ is the Lindblad operator that accounts for the decay processes in the atom.

For the six-level system shown in Fig.~\ref{multilevel}, the Hamiltonian can be expressed as:
\begin{equation}
\begin{footnotesize}
H=\frac{\hbar}{2}\left[\begin{array}{cccccc}
0 & \Omega_p & 0 & 0&0&0\\
\Omega_p & A & \Omega_c & 0&0&0\\
0 & \Omega_c & B & \Omega_{SIG34}&0&0\\
0 & 0 & \Omega_{SIG34} & C &\Omega_{ARRT}&\Omega_{SIG46}\\
0 & 0 &0& \Omega_{ARRT} & D &0\\
0 & 0 &0& \Omega_{SIG46} & 0 &E\\
\end{array}
\right]\,\, ,
\end{footnotesize}
\label{H4}
\end{equation}
where $\Omega_p$, $\Omega_c$, and $\Omega_{SIG}$ are the Rabi frequencies of the probe laser, coupling laser, and ARRT field coupled states, respectively. Also,
\begin{equation}
\begin{array}{rcl}
A&=&2\Delta_p \\
B&=&-2(\Delta_p+\Delta_c)\\
C&=&-2(\Delta_p+\Delta_c+\Delta_{SIG34})\\
D&=&-2(\Delta_p+\Delta_c+\Delta_{SIG34}+\Delta_{ARRT})\\
E&=&-2(\Delta_p+\Delta_c+\Delta_{SIG34}+\Delta_{SIG46}) ~~ ,\\
\end{array}
\end{equation}
where $\Delta_p$, $\Delta_c$, and $\Delta_{ARRT}$ are the detunings of the probe laser, couple laser, and the ARRT field, respectively, defined as
\begin{equation}
\Delta_{p,c,ARRT}=\omega_{p,c,ARRT}-\omega_{12,23,45} \,\,\, ,
\label{detuningeq}
\end{equation}
where $\omega_{12,23,45}$ are the on-resonance angular frequencies of transitions $\ket{2}$-$\ket{3}$, and $\ket{4}$-$\ket{5}$ for the probe, coupling, and ARRT fields, respectively, and $\omega_{p,c,T}$ are the angular frequencies of the probe, coupling, and ARRT fields, respectively. The SIG field detuning is defined as
\begin{equation}
    \Delta_{SIG34} = \Delta_{SIG} = \omega_{SIG} - \omega_{34}
\end{equation}
for the $\ket{3}-\ket{4}$ transition, and detuning of the additional $\ket{4}-\ket{6}$ transition is defined as
\begin{align}
    &\Delta_{SIG46} = \omega_{SIG} - \omega_{46} \nonumber\\
    &= 2\pi \cdot \left[12.979~\mathrm{GHz}-23.888~\mathrm{GHz}\right] + \Delta_{SIG}.
\end{align}

\begin{figure}[htbp]
\includegraphics[width = \columnwidth]{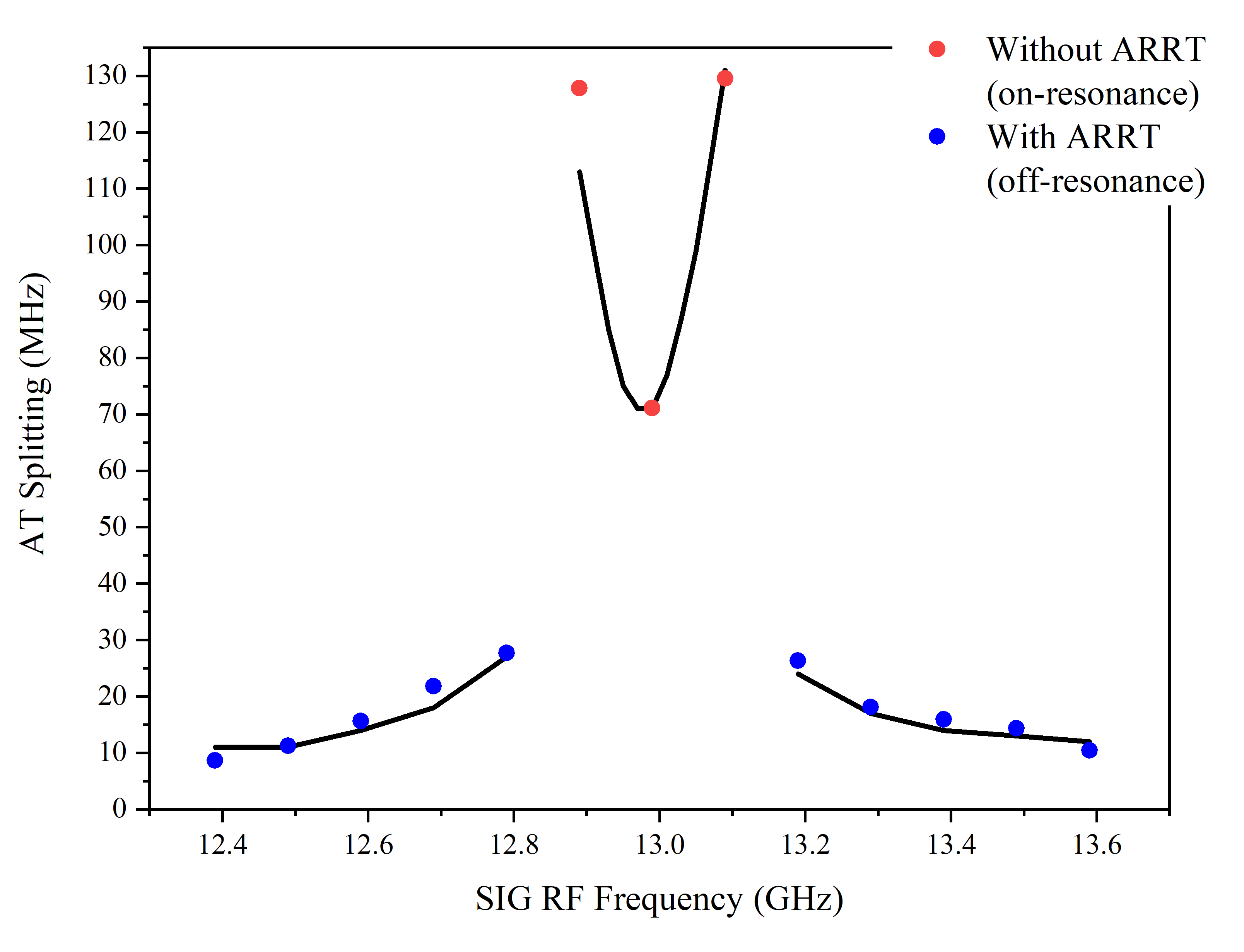}
\caption{\label{fig:sim1} Comparison of simulations to experimental data in Fig.~\ref{fig:alldetuning}, from the SIG transition $68~S_{1/2} - 67~P_{3/2}$ using an ARRT field near $67~P_{3/2} - 65~D_{5/2}$. The AT splittings are predicted well by theory.}
\end{figure}

\begin{table*}[htbp]
\begin{minipage}{0.75\textwidth}
\begin{equation}
\begin{footnotesize}
{\cal{L}}=\left[\begin{array}{cccccc}
\Gamma_2 \rho_{22} & -\gamma_{12}\rho_{12} & -\gamma_{13}\rho_{13} & -\gamma_{14}\rho_{14}& -\gamma_{15}\rho_{15}  & -\gamma_{16}\rho_{16} \\

-\gamma_{21}\rho_{21} & \Gamma_3 \rho_{33}-\Gamma_2 \rho_{22} & -\gamma_{23}\rho_{23} & -\gamma_{24}\rho_{24}& -\gamma_{25}\rho_{25}  & -\gamma_{26}\rho_{26}\\

-\gamma_{31}\rho_{31} & -\gamma_{32}\rho_{32} & -\Gamma_3 \rho_{33} & -\gamma_{34}\rho_{34}& -\gamma_{35}\rho_{35}  & -\gamma_{36}\rho_{36}\\

-\gamma_{41}\rho_{41} & -\gamma_{42}\rho_{42} & -\gamma_{43}\rho_{43} &  \Gamma_3 \rho_{33}-\Gamma_4 \rho_{44}& -\gamma_{45}\rho_{45}  & -\gamma_{46}\rho_{46}\\

-\gamma_{51}\rho_{51} & -\gamma_{52}\rho_{52} & -\gamma_{53}\rho_{53} & -\gamma_{45}\rho_{45}  &  \Gamma_4 \rho_{44}-\Gamma_5 \rho_{55}& -\gamma_{56}\rho_{56}\\

-\gamma_{61}\rho_{61} & -\gamma_{62}\rho_{62} & -\gamma_{63}\rho_{63} & -\gamma_{65}\rho_{65} & -\gamma_{56}\rho_{56} &  \Gamma_4 \rho_{44}-\Gamma_6 \rho_{66}\\

\end{array}
\right]
\end{footnotesize}
\label{L4}
\end{equation}
\end{minipage}
\end{table*}

In our experiments, $\Delta_p=0~\mathrm{rad/s}$ and $\Delta_c$ is scanned, while $\Delta_{SIG}$ and $\Delta_{ARRT}$ are varied. The optical Rabi frequencies are $\Omega_p/2\pi=9.48~\mathrm{MHz}$ and $\Omega_c/2\pi=0.564~\mathrm{MHz}$ for the probe and coupling lasers, respectively. The signal RF field Rabi frequency is generally held fixed at $\Omega_{SIG}/2\pi=100~\mathrm{MHz}$ or thereabouts, while the ARRT field Rabi frequency $\Omega_{T}$ is varied.

For the six-level system, the ${\cal{L}}$ matrix is given in eq.~(\ref{L4}), where $\gamma_{ij}=(\Gamma_i+\Gamma_j)/2$ and $\Gamma_{i, j}$ are the transition decay rates. Since the purpose of the present study is to explore the intrinsic limitations of Rydberg-EIT field sensing in vapor cells, no collision terms or dephasing terms are added. While Rydberg-atom collisions, Penning ionization, and ion electric fields can, in principle, cause dephasing, such effects can, for instance, be alleviated by reducing the beam intensities, lowering the vapor pressure, or limiting the atom-field interaction time. In this analysis we set,
$\Gamma_1=0$, $\Gamma_2=2\pi\times$(6~{\rm MHz}),
$\Gamma_3=2\pi\times$(3~{\rm kHz}), $\Gamma_{4,5,6}=2\pi\times$(2~{\rm kHz}).
Note, $\Gamma_{2}$ is for the D2 line in $^{133}$Cs \cite{SteckCsData}, and $\Gamma_{3}$, $\Gamma_{4,5,6}$, are typical Rydberg decay rates.

\begin{figure}
    \centering
    \includegraphics[width=\columnwidth]{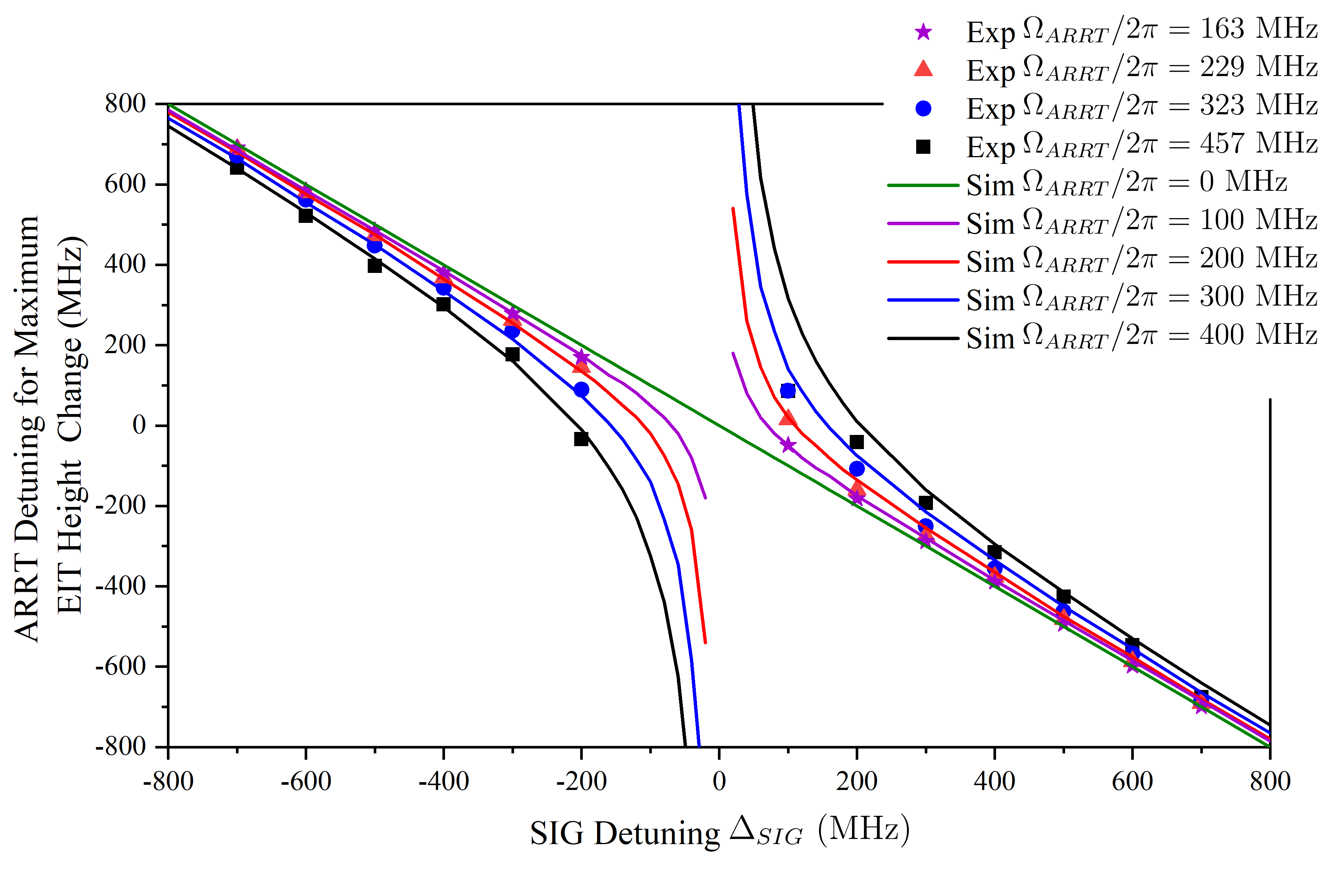}
    \caption{Repeat of Fig.~\ref{fig:pow3} with more $\Omega_{T}$ measurements and compared to the model calculations. The experimental measurements show an asymmetry with the sign of $\Delta_{SIG}$, suggestive of a frequency shifting that the model does not capture, whereas the model does demonstrate the divergent trend as the SIG approaches the on-resonant condition (where the ARRT field is no longer beneficial).}
    \label{fig:model_pow3}
\end{figure}

We numerically solve these equations to find the steady-state solution for $\rho_{21}$ for various values of $\Omega_{SIG}$, $\Omega_{ARRT}$, $\Delta_{SIG}$, and $\Delta_{ARRT}$. This is done by forming a matrix with the system of equations for $\dot{\rho}_{ij}=0$. The null-space of the resulting system matrix is the steady-state solution.  The steady-state solution for $\rho_{21}$ is then Doppler averaged\cite{berman2011book}
\begin{equation}
\rho_{21_D}=\frac{1}{\sqrt{\pi}\,\, u}\int_{-3u}^{3u}\rho_{21}\left(\Delta'_p,\Delta'_c\right)\,\,e^{\frac{-v^2}{u^2}}\,\,dv\,\,\, ,
\label{doppler}
\end{equation}
where $u=\sqrt{2k_B T/m}$ and $m$ is the mass of the atom. We use the case where the probe and coupling laser are counter-propagating. Thus, the frequency seen by the atom moving toward the probe beam is upshifted by $2\pi v/\lambda_p$ (where $v$ is the velocity of the atoms), while the frequency of the coupling beam seen by the same atom is downshifted by $2\pi v/\lambda_c$.  The probe and coupling beam detuning is modified by the following
\begin{equation}
\Delta'_p=\Delta_p-\frac{2\pi}{\lambda_p}v \,\,\,{\rm and}\,\,\,
\Delta'_c=\Delta_c+\frac{2\pi}{\lambda_c}v \,\,\, .
\label{doppler2}
\end{equation}

Figure~\ref{fig:sim1} shows a comparison of numerically generated results from the theoretical model and experimental data. A duplication of the $68~S_{1/2} \rightarrow 67~P_{3/2}$ state transition measurements from Fig.~\ref{several-n}, this figure shows that the model reproduces the general effect of having the ARRT field on for off-resonant detection of SIG. Similarly, Fig.~\ref{fig:alldetuning} compares the model with measurements of AT splitting over a wide range of SIG detunings for a set of ARRT detunings with good agreement of the model to the experimental measurements. Small differences between the model and experimental results in this figure may be due to variations in the SIG or ARRT Rabi frequencies over the full range of SIG and ARRT frequencies. We also compare the measurements reported in Fig.~\ref{fig:pow3} to the model in Fig.~\ref{fig:model_pow3} and again see that the model recreates the measured trends. However, where an asymmetry was measured in the experiments (likely due to an AC stark shift caused by the tuning field) the model does not reproduce this effect.

\begin{figure}
    \centering
    \includegraphics[width=\columnwidth]{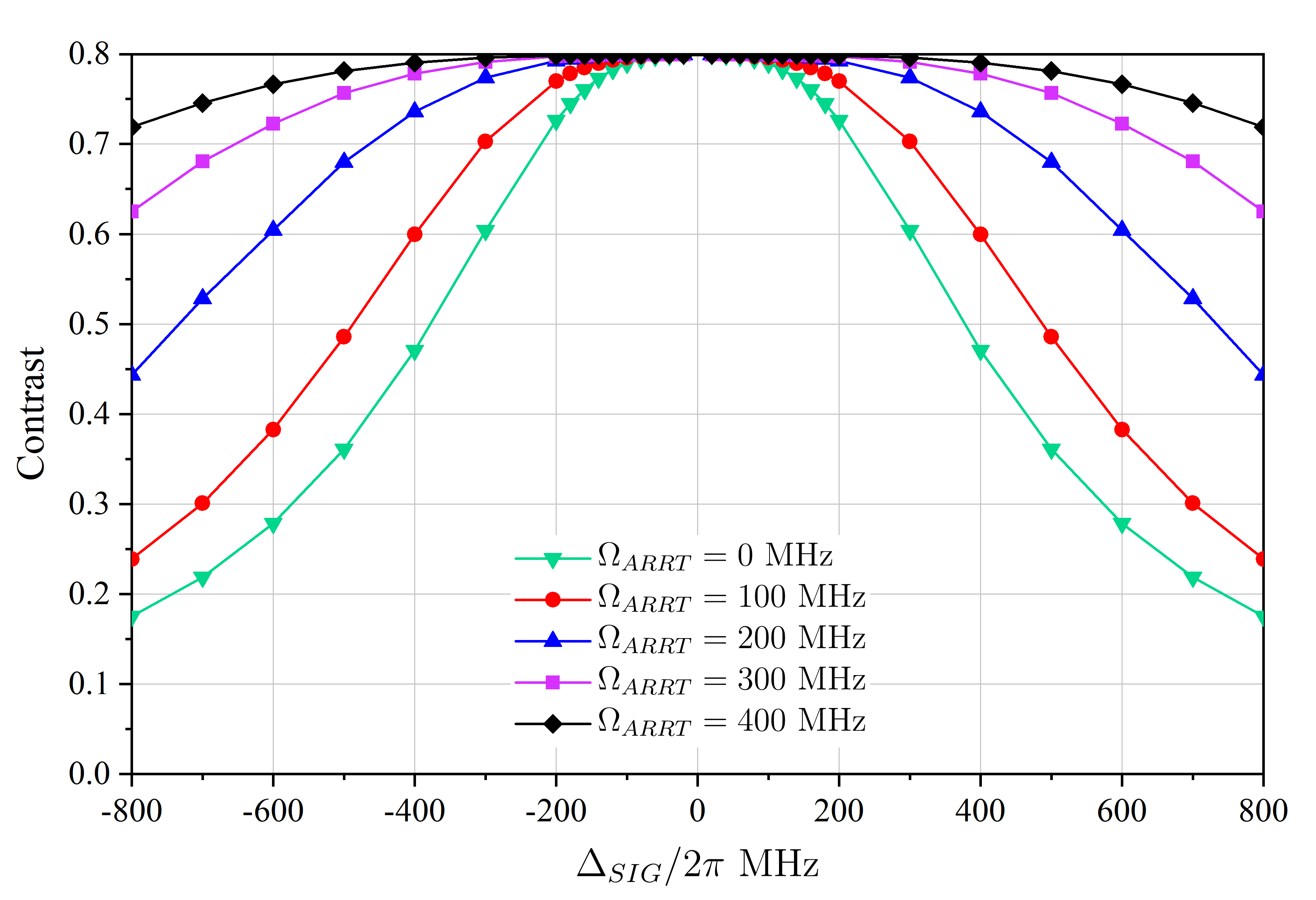}
    \caption{Maximum EIT height change (from Fig.~\ref{fig:model_pow3}) reported as contrast (Eq.~\ref{eq:contrast}) for a set of SIG and ARRT Rabi frequencies ($\Omega_{SIG}$ and $\Omega_{ARRT}$, respectively) over a wide range of SIG detunings ($\Delta_{SIG}$). Increased contrast relates to an increase in sensitivity to the detuned SIG field.}
    \label{fig:contrast}
\end{figure}

A key benefit of having a theoretical model that reproduces experimental results of this kind is the ability to quickly study a wider parameter space than is feasible purely with experimental measurements. For example, we take the model results of Fig.~\ref{fig:model_pow3} and calculate a contrast parameter for the change in EIT height given by
\begin{equation}\label{eq:contrast}
    C = \frac{|T_{RF}-T_{0}|}{T_0},
\end{equation}
where $T_{0}$ is the probe laser transmittance ($P/P_0$ from Eq.~\ref{intensity}) for $\Delta_p=\Delta_c=0~\mathrm{rad/s}$ with no RF fields ($\Omega_{SIG}=\Omega_{T}=0~\mathrm{rad/s}$) and $T_{RF}$ is the same probe laser transmittance but with SIG and ARRT fields turned on. The sensitivity of a measurement of SIG is directly related to this contrast parameter such that high EIT contrast relates to a high sensitivity. Figure~\ref{fig:contrast} gives the modeled calculations of contrast for the set of ARRT Rabi frequencies over a $\pm800~\mathrm{MHz}$ range of SIG detunings. We note that the model considers only the states defined in Fig.~\ref{multilevel}, so SIG detunings beyond this range are not well predicted by the model. It is clear from Fig.~\ref{fig:contrast} that increasing the power (i.e., the Rabi frequency) of the ARRT field results in significant increases in contrast (or measurement sensitivity) for SIG fields that are highly detuned from resonance. 

\begin{figure}
    \centering
    \includegraphics[width=\columnwidth]{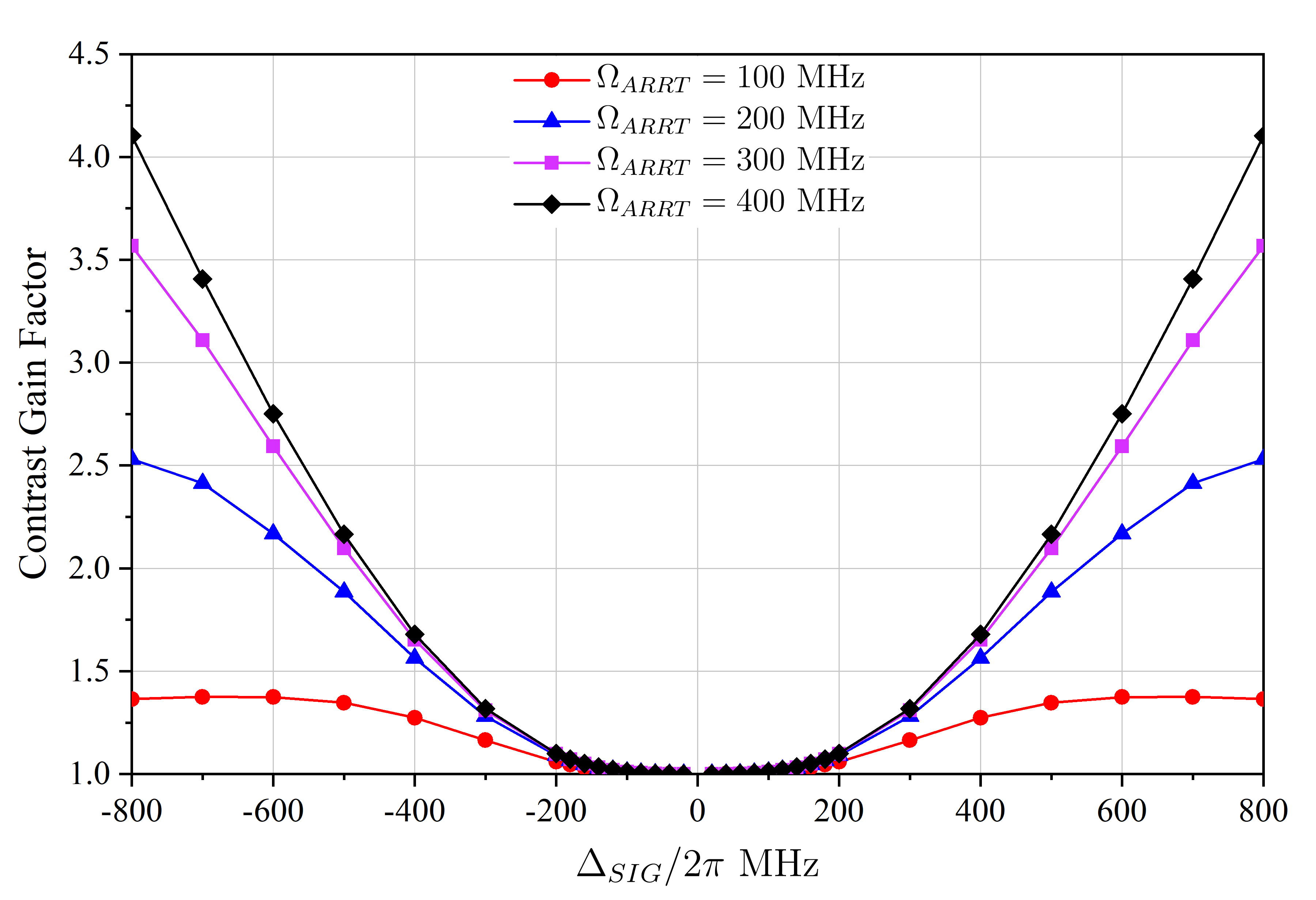}
    \caption{Numerically modeled contrast gain factor (Eq.~\ref{eq:contrast_gain}) for a set of ARRT field Rabi frequencies ($\Omega_T$) over a wide range of SIG field detunings ($\Delta_{SIG}$) showing increased sensitivity gain for strong ARRT fields and highly detuned SIG fields.}
    \label{fig:contrast-gain}
\end{figure}

We can quantify the improvement obtained in the measurement by the incorporation of this ARRT field by calculating the contrast gain factor,
\begin{equation}\label{eq:contrast_gain}
    G = \frac{C_{tuner}}{C_{0}},
\end{equation}
or the ratio of calculated contrast values with and without an ARRT field (no ARRT field is modeled by the orange curve in Fig.~\ref{fig:model_pow3} and Fig.~\ref{fig:contrast} where $\Omega_{T}/2\pi=0~\mathrm{MHz}$). These contrast gains calculated by our model are given in Fig.~\ref{fig:contrast-gain} for the same set of $\Omega_{ARRT}$ and SIG detunings. With a strong ARRT field, the EIT height change, or the contrast, is increased by as much as four times for highly detuned SIG fields.

\section{Conclusion}

In this paper, we demonstrate a five-level atomic scheme that allows for the resonant detection of an RF signal by Rydberg EIT over a continuous frequency range. The main Rydberg transition frequency is tuned by the presence of an adjacent Rydberg resonance tuning field. In this five-level scheme, we show continuous frequency detection can be obtained by either varying the frequency or the power level of the ARRT field. Notably, this technique allows for the detection of a signal RF field over a frequency range that covers from one Rydberg state with principal quantum number $n$ to the next $n+1$ state. The data in this paper show that the Autler-Townes tuning of Rydberg resonances allows for complete spectral coverage;  the dressed states may be shifted by an amount corresponding to a change of the principal quantum number by 1 without significant degradation of the electric field sensitivity.  

\newpage

\bibliography{myrefs}

\end{document}